\documentclass[proceedings, preprint]{rmaa}

\usepackage{paralist}

\usepackage{psfrag,color}

\newcommand{\vc}[1]{\textbf{\em #1}}

\newcommand{\pder}[2]{\frac{\partial #1}{\partial #2}}

\SetYear{2008}
\SetConfTitle{Magnetic Fields in the Universe II: From Laboratory and Stars to the Primordial Universe}

\title{Numerical Studies of Weakly Stochastic Magnetic Reconnection}

\author{
  G. Kowal,\altaffilmark{1,2}
  A. Lazarian,\altaffilmark{1}
  E. T. Vishniac\altaffilmark{3}
  and K. Otmianowska-Mazur\altaffilmark{2}}

\altaffiltext{1}{Department of Astronomy, UW-Madison,
  475 North Charter Street, Madison, WI 53719,
  USA (kowal@astro.wisc.edu, lazarian@astro.wisc.edu).}

\altaffiltext{2}{Astronomical Observatory, Jagiellonian University,
  ul. Orla 171, 30-244 Krak\'ow, Poland (kowal@oa.uj.edu.pl, otmian@oa.uj.edu.pl).}

\altaffiltext{3}{Department of Physics and Astronomy, McMaster University,
  1280 Main Street West, Hamilton, ON L8S 4M1, Canada (ethan@mcmaster.ca).}

\shortauthor{Kowal et al.}
\shorttitle{Weakly Stochastic Magnetic Reconnection}

\listofauthors{G. Kowal, A. Lazarian, E. T. Vishniac, K. Otmianowska-Mazur}
\indexauthor{Kowal, G.}
\indexauthor{Lazarian, A.}
\indexauthor{Vishniac, E. T.}
\indexauthor{Otmianowska-Mazur, K.}

\abstract{We study the effects of turbulence on magnetic reconnection using
three-dimensional numerical simulations. This is the first attempt to test a
model of fast magnetic reconnection proposed by Lazarian \& Vishniac (1999),
which assumes the presence of weak, small-scale magnetic field structure near
the current sheet. This affects the rate of reconnection by reducing the
transverse scale for reconnection flows and by allowing many independent flux
reconnection events to occur simultaneously. We performed a number of
simulations to test the dependencies of the reconnection speed, defined as the
ratio of the inflow velocity to the Alfv\'en speed, on the turbulence power, the
injection scale and resistivity. Our results show that turbulence significantly
affects the topology of magnetic field near the diffusion region and increases
the thickness of the outflow region. We confirm the predictions of the Lazarian
\& Vishniac model. In particular, we report the growth of the reconnection speed
proportional to $\sim V_l^2$, where $V_l$ is the amplitude of velocity at the
injection scale. It depends on the injection scale $l_{inj}$ as $\sim
(l_{inj}/L)^{2/3}$, where $L$ is the size of the system, which is somewhat
faster but still roughly consistent with the theoretical expectations. We also
show that for 3D reconnection the Ohmic resistivity is important in the local
reconnection events only, and the global reconnection rate in the presence of
turbulence does not depend on it.}

\addkeyword{galaxies: magnetic fields}
\addkeyword{physical processes: MHD}
\addkeyword{physical processes: turbulence}
\addkeyword{methods: numerical}

\begin{document}
% Typeset article header
\maketitle

%%=============================================================================
%%
%%
\section{Introduction}
\label{sec:intro}

Magnetic fields play a key role in the astrophysical processes such as star
formation, the transport and acceleration of cosmic rays, accretion disks, solar
phenomena, etc. Typical magnetic diffusion is very slow on astrophysical scales,
so the sufficient approximation of the evolution of magnetic field is its
advection with the flow, i.e. the magnetic field is ''frozen-in'' and moves
together with the medium \citep[see][]{moffat78}.

Reconnection is a fundamental process that describes how bundles of magnetic
field lines can pass through each other. Different approaches to the problem of
reconnection are discussed in the companion paper by \citet{lazarian08}. In what
follows we concentrate on testing the model of reconnection for a weakly
stochastic field proposed by \citet[][LV99 henceforth]{lazarian99}. They argued
that reconnection speed is equal to the upper limit imposed by large-scale field
line diffusion, expressed by
\begin{equation}
V_{rec} = V_A \min \left[ \left( \frac{L}{l} \right)^{1/2}, \left( \frac{l}{L} \right)^{1/2} \right] \left( \frac{V_l}{V_A} \right)^{1/2},
\label{eq:constraint}
\end{equation}
where $V_A$ is the Alfv\'en speed, $L$ is the size of the system, $l$ is the
injection scale, and $V_l$ is the velocity amplitude at the injection scale. In
this relation, the reconnection speed is determined by the characteristics of
turbulence, namely, its strength and injection scale. Most importantly, there is
no explicit dependence on the Ohmic resistivity.

The numerical testing of reconnection models is far from trivial. While most of
the reconnection work \citep[see][]{priest00} is performed in 2D, the LV99 model is intrinsically 3D.

In \S\ref{sec:model} we describe our numerical model, in \S\ref{sec:rec_rate} we
describe the reconnection rate we use, in \S\ref{sec:results} and
\S\ref{sec:discussion} we describe and discuss the results we obtained, and in
\S\ref{sec:summary} we summarize the paper drawing the main conclusions.

%%=============================================================================
%%
%%
\section{Numerical Modeling of LV99 Reconnection}
\label{sec:model}

%%-----------------------------------------------------------------------------
%%
\subsection{Governing Equations}
\label{ssec:equations}

We use a higher-order shock-capturing Godunov-type scheme based on the
essentially non oscillatory (ENO) spacial reconstruction and Runge-Kutta (RK)
time integration \citep[see][e.g.]{delzanna03} to solve isothermal non-ideal MHD
equations,
\begin{eqnarray}
 \pder{\rho}{t} + \nabla \cdot \left( \rho \vc{v} \right) & = & 0, \label{eq:mass} \\
 \pder{\rho \vc{v}}{t} + \nabla \cdot \left[ \rho \vc{v} \vc{v} + p_T I - \frac{\vc{B} \vc{B}}{4 \pi} \right] & = & \vc{f}, \label{eq:momentum} \\
 \pder{\vc{A}}{t} + \vc{E} & = & 0 \label{eq:induction},
\end{eqnarray}
where $\rho$ and $\vc{v}$ are plasma density and velocity, respectively,
$\vc{A}$ is vector potential, $\vc{E} = - \vc{v} \times \vc{B} + \eta \, \vc{j}$
is electric field, $\vc{B} \equiv \nabla \times \vc{A}$ is magnetic field,
$\vc{j} = \nabla \times \vc{B}$ is current density, $p_T = a^2 \rho + B^2 / 8
\pi$ is the total pressure, $a$ is the isothermal speed of sound, $\eta$ is
resistivity coefficient, and $\vc{f}$ represents the forcing term.

We incorporated the field interpolated constrained transport (CT) scheme
\citep[see][]{toth00} in to the integration of the induction equation to
maintain the $\nabla \cdot \vc{B} = 0$ constraint numerically.

Some selected simulations that we perform include anomalous resistivity modeled
as
\begin{equation}
\eta = \eta_u + \eta_a \left( \frac{| \vc{j} |}{j_\mathrm{crit}} - 1 \right) H \left( \frac{| \vc{j} |}{j_\mathrm{crit}} \right),
\end{equation}
where $\eta_u$ and $\eta_a$ describe uniform and anomalous resistivity
coefficients, respectively, $j_{crit}$ is the critical level of the absolute
value of current density $\vc{j}$ above which the anomalous effects start to
work, and $H$ is a step function. For most of our simulations $\eta_a = 0$,
however.

%%-----------------------------------------------------------------------------
%%
\subsection{Initial Conditions and Parameters}
\label{ssec:initial}

Our initial magnetic field is a Harris current sheet of the form $B_x = B_{x0}
\tanh (y/\theta)$ initialized using the magnetic vector potential $A_z = \ln |
\cosh(y / \theta) |$. In addition, we use a uniform shear component $B_z =
B_{z0} = \mathrm{const}$. The initial setup is completed by setting the density
profile from the condition of the uniform total pressure $p_T(t=0) =
\mathrm{const}$ and setting the initial velocity to zero everywhere.

In order to initiate the magnetic reconnection we add a small initial
perturbation of vector potential $\delta A_z = B_{x0} \cos(2 \pi x)
\exp[-(y/d)^2]$ to the initial configuration of $A_z(t=0)$. The parameter $d$
describes the thickness of the perturbed region.

Numerical model of the LV99 reconnection is evolved in a box with open boundary
conditions which we describe in the next sub-section. The box has sizes $L_x =
L_z = 1$ and $L_y=2$ with the resolution 256x512x256. It is extended in
Y-direction in order to move the inflow boundaries far from the injection
region. This minimizes the influence of the injected turbulence on the inflow.

Initially, we set the strength of anti-parallel magnetic field component to 1
and we vary the shear component $B_z$ between 0.0 and 1.0. The speed of sound is
set to 4. In order to study the resistivity dependence on the reconnection we
vary the resistivity coefficient $\eta$ between values $0.5\cdot10^{-4}$ and
$2\cdot10^{-3}$ which are expressed in dimensionless units. This means that the
velocity is expressed in units of Alfv\'en speed and time in units of Alfv\'en
time $t_A = L / V_A$, where $L$ is the size of the box.

%%-----------------------------------------------------------------------------
%%
\subsection{Boundary Conditions}
\label{ssec:boundaries}

The boundary conditions are set for the fluid quantities and the magnetic vector
potential separately. For density and velocity we solve a wave equation with the
speed of propagation equal to the maximum linear speed which is the fast
magnetosonic one and its sign corresponding to the outgoing wave. This assumption
guarantees that all waves generated in the system are free to leave the box
without significant reflections. Moreover, the waves propagate through the
boundary with the maximum speed, reducing the chance for interaction with
incoming waves.

For the vector potential, its perpendicular components to the boundary are
obtained using the first order extrapolation, while the normal component has
zero normal derivative. This guarantees that the normal derivative of the
magnetic field components is zero, which reduces the influence of the magnetic
field at the boundary on the plasma flow.

This type of boundary conditions represents a mixed inflow/outflow boundaries,
which are adjusting during the evolution of the system. It means that we do not
set fixed values of quantities and do not drive the flow at the boundaries in
order to achieve a stationary reconnection.

%%-----------------------------------------------------------------------------
%%
\subsection{Method of Driving Turbulence}
\label{ssec:turbulence}

In our model we drive turbulence using a method described by \citet{alvelius99}.
The forcing is implemented in spectral space where it is concentrated with a
Gaussian profile around a wave vector corresponding to the injection scale
$l_{inj}$. Since we can control the scale of injection, the energy input is
introduced into the flow at arbitrary scale. The randomness in the time makes
the force neutral in the sense that it does not directly correlate with any of
the time scales of the turbulent flow and it also makes the power input
determined solely by the force-force correlation. This means that it is possible
to generate different states of turbulence, such as axisymmetric turbulence,
where the degree of anisotropy of the forcing can be chosen {\em a priori}
through the forcing parameters. In the present paper we limit our studies to the
isotropic forcing only. The total amount of power input from the forcing can be
set to balance a desired dissipation at a statistically stationary state. In
order to get the contribution to the input power in the discrete equations from
the force-force correlation only, the force is determined so that the
velocity-force correlations vanish for each Fourier mode.

On the right hand side of Equation~(\ref{eq:momentum}), the forcing is
represented by a function $\vc{f} = \rho \vc{a}$, where $\rho$ is the local
density and $\vc{a}$ is a random acceleration calculated using the method
described above.

The driving is completely solenoidal, which means that it does not produce
density fluctuations. Density fluctuations results from the wave interaction
generated during the evolution of the system. Nevertheless, in our models we set
large values of the speed of sound approaching nearly incompressible regime of
turbulence.

%%=============================================================================
%%
%%
\section{Reconnection Rate Measure}
\label{sec:rec_rate}

We measure the reconnection rate by averaging the inflow velocity $V_{in}$
divided by the Alfv\'en speed $V_A$ over the inflow boundaries. In this way our
definition of the reconnection rate is
\begin{equation}
V_{rec} = \langle V_{in} / V_A \rangle_{S} = \int_{y=y_{min},y_{max}}{ \frac{\vec{V}}{V_A}  \cdot d\vec{S}},
\end{equation}
where $S$ defines the XZ planes of the inflow boundaries.

%%=============================================================================
%%
%%
\section{Results}
\label{sec:results}

In this section we describe the results obtained from the three dimensional
simulations of the magnetic reconnection in the presence of turbulence. First,
we investigate the Sweet-Parker reconnection, a stage before we inject
turbulence. A full understanding of this stage is required before we perform further
analysis of reconnection in the presence of turbulence.

%%=============================================================================
%%
\subsection{Sweet-Parker Reconnection}
\label{ssec:sweet-parker}

As we described in \S\ref{ssec:initial}, Sweet-Parker reconnection develops
in our models as a result of an initial vector potential perturbation. In order
to reliably study the influence of turbulence on the evolution of such systems,
we need to reach the stationary Sweet-Parker reconnection before we start
injecting turbulence.

\begin{figure}[t]
\includegraphics[width=\columnwidth]{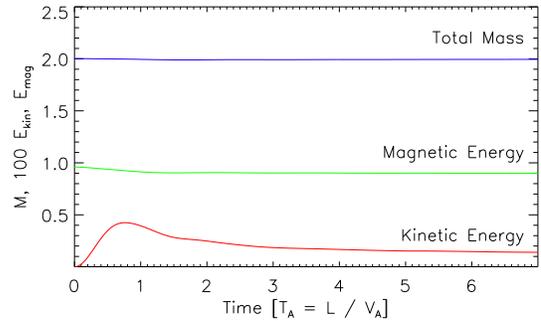}
\caption{Evolution of total mass $M$ and kinetic and magnetic energies,
$E_{kin}$ and $E_{mag}$ respectively. The kinetic energy $E_{kin}$ has been
amplified by a factor of 100 to visualize its evolution more clearly. The
resistivity in this model is $\eta=10^{-3}$ and the shear component of magnetic
field $B_z = 0.1$. \label{fig:conservation}}
\end{figure}
Figure~\ref{fig:conservation} shows the evolution of total mass, kinetic and
magnetic energies until the moment at which we start injecting the turbulence
(i.e. $t=7$). All shown quantities, after some initial adaptation, reach steady,
almost constant values. The near zero time derivatives of total mass, kinetic
and magnetic energies guarantee the stationarity of the system. We remind the
reader, that the system evolves in the presence of open boundary conditions,
which could violate the conservation of mass and total energy. As we see, the
conservation of these quantities is well satisfied during the Sweet-Parker stage
in our models.

\begin{figure}[h]
\includegraphics[width=\columnwidth]{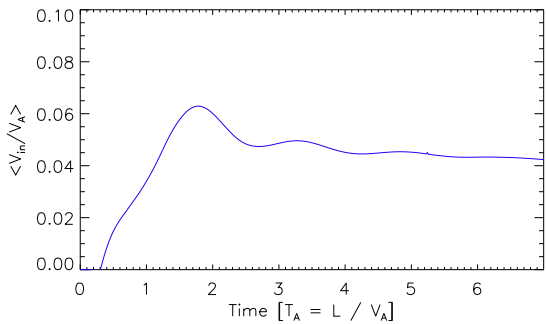}
\caption{Evolution of the reconnection rate $\langle V_{in}/V_A \rangle$ for the
same model as in Fig.~\ref{fig:conservation}. The reconnection rate grows
initially until it reaches the stationary solution. \label{fig:rate_sp}}
\end{figure}
The reconnection rate, shown in Figure~\ref{fig:rate_sp}, also confirms that the
evolution reaches a stationary state. Initially, the reconnection rate $\langle
V_{in}/V_A \rangle$ grows until time $t \approx 1.8$, when it reaches the
maximum value of $\approx 0.06$. Later on, it drops a bit approaching a
value of $0.04$. During the last period of about 3 Alfv\'en time units, the
change of the reconnection rate is very small. We assume that these conditions
guarantee a nearly steady state evolution of the system, so at this point we are
ready to introduce turbulence.

In Figure~\ref{fig:top_sp} we present the velocity and magnetic field
configuration of the steady state. In the left panel of the figure we show the
topology of velocity field as textures. The brightness of texture corresponds to
the amplitude of velocity. The texture itself shows the direction of the field
lines. The topology of the velocity field is mainly characterized by strong outflow
regions along the mid-plane. The outflow is produced by the constant
reconnection process at the diffusion region near the center and the ejection of the
reconnected magnetic flux through the left and right boundaries. The system is
in a steady state when the flux which reconnects is counterbalanced by the
incoming fresh flux. The inflow is much slower then the outflow,
but still its direction can be recognized from the texture shown in the left
plot of Figure~\ref{fig:top_sp}.

The topology of the magnetic field is presented in the middle panel of
Figure~\ref{fig:top_sp}. We recognize the anti-parallel configuration of the
field lines with the uniform strength out of the mid-plane. Near the mid-plane,
the horizontal magnetic lines are reconnected generating the Y-component, which
is ejected by the strong outflow. In addition, we show the absolute value of
current density in the right panel of Figure~\ref{fig:top_sp}. We see very
elongated diffusion region in the middle of the box, where reconnection
takes place. The maximum value of $|\vec{j}|$ does not exceed a value of
25.

%%=============================================================================
%%
\subsection{Effects of Turbulence}
\label{ssec:turbulent}

The essential part of our studies covers the effects of turbulence on
reconnection. Our goal is to achieve a stationary state of Sweet-Parker
reconnection, which is described in the previous subsection, and then introduce
turbulence at a given injection scale $l_{inj}$, gradually increasing its
strength to the desired amplitude corresponding to the turbulent power
$P_{inj}$. We inject turbulence in the region surrounding the mid plane and
extending to the distance of around one quarter of the size of the box. The
transition period during which we increase the strength of turbulence, has
length of one Alfv\'enic time and starts at $t=7$. It means that from $t=8$ we
inject turbulence at maximum power $P_{inj}$.

\begin{figure*}
\center
\includegraphics[width=0.6\columnwidth]{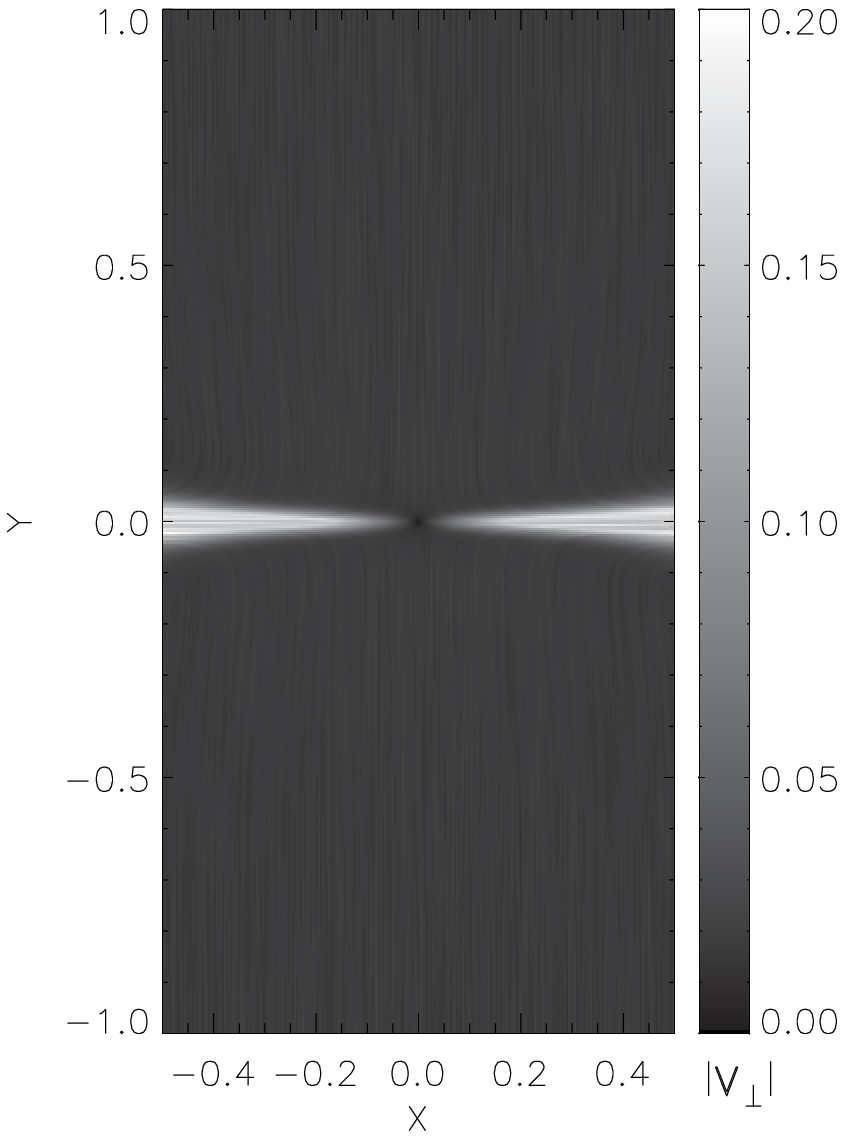}
\includegraphics[width=0.6\columnwidth]{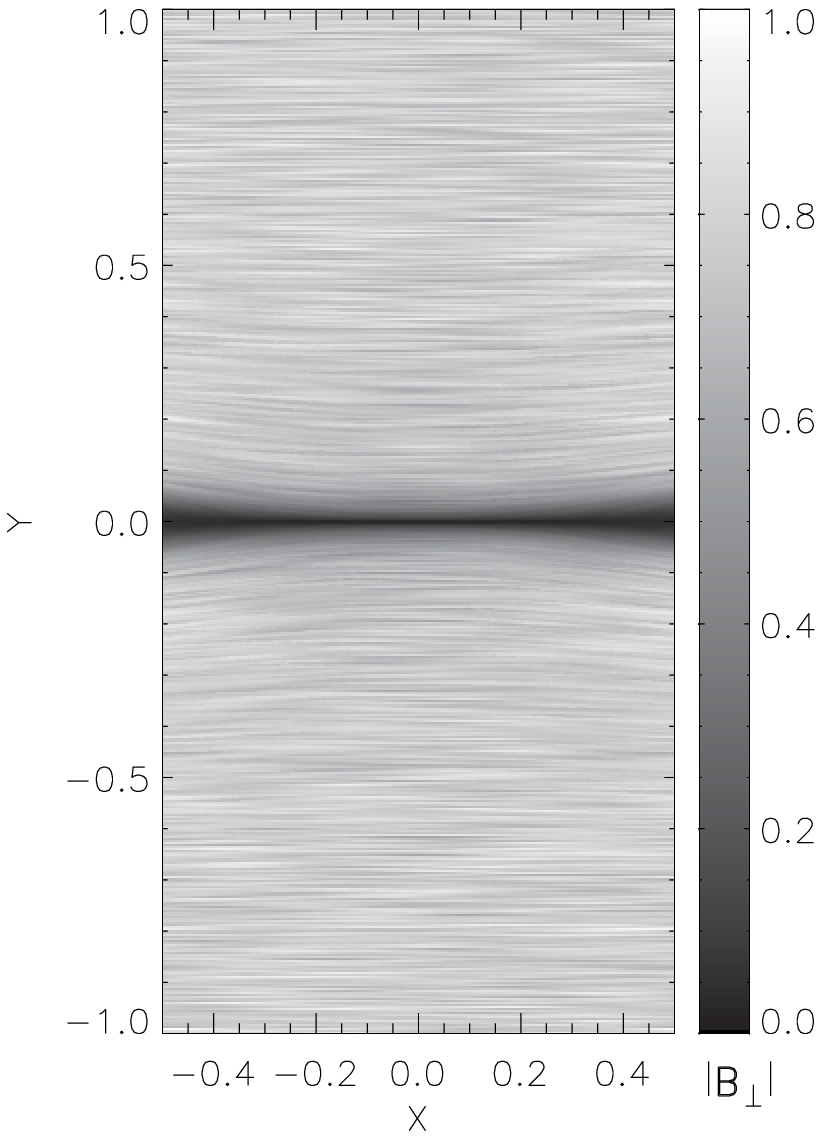}
\includegraphics[width=0.6\columnwidth]{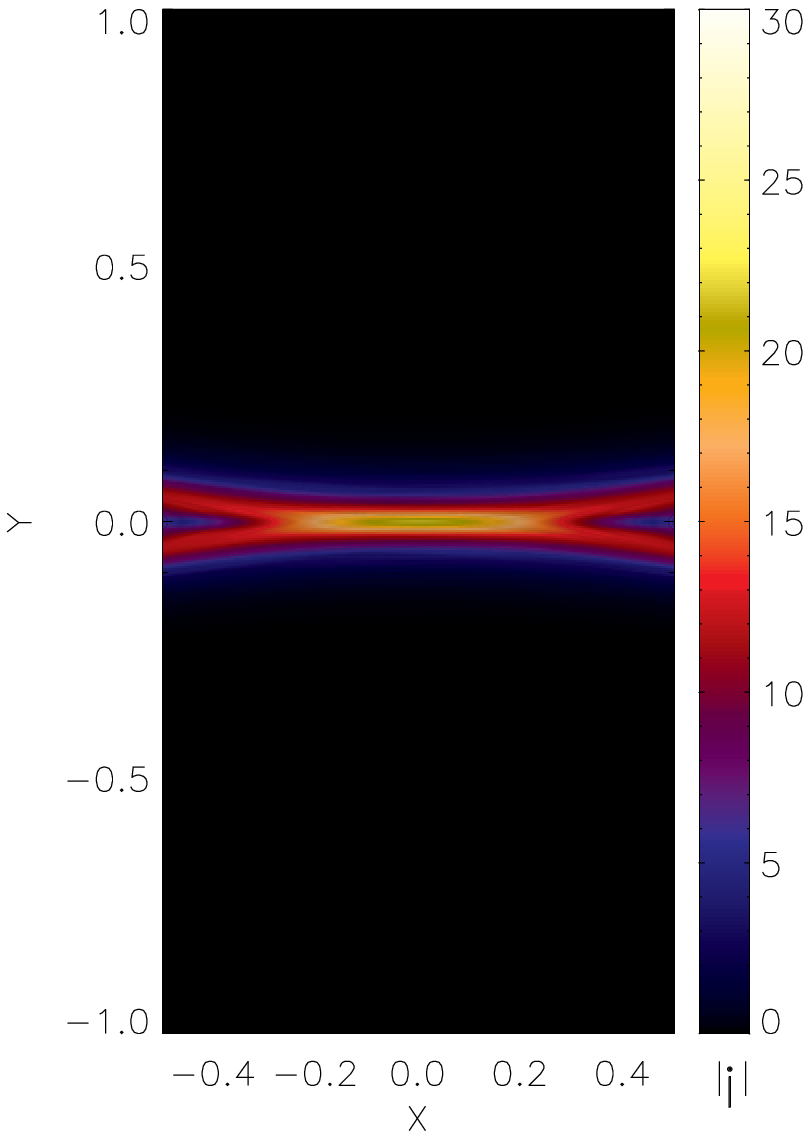}
\caption{Topology and strength of velocity field (left panel) and magnetic field
(middle panel) during the Sweet-Parker reconnection at $t=7$. In the right
panel we show the absolute value of current density $\vec{j}$. The images show
the XY-cut through the domain at $Z=0$. \label{fig:top_sp}}
\end{figure*}
\begin{figure*}
\center
\includegraphics[width=0.6\columnwidth]{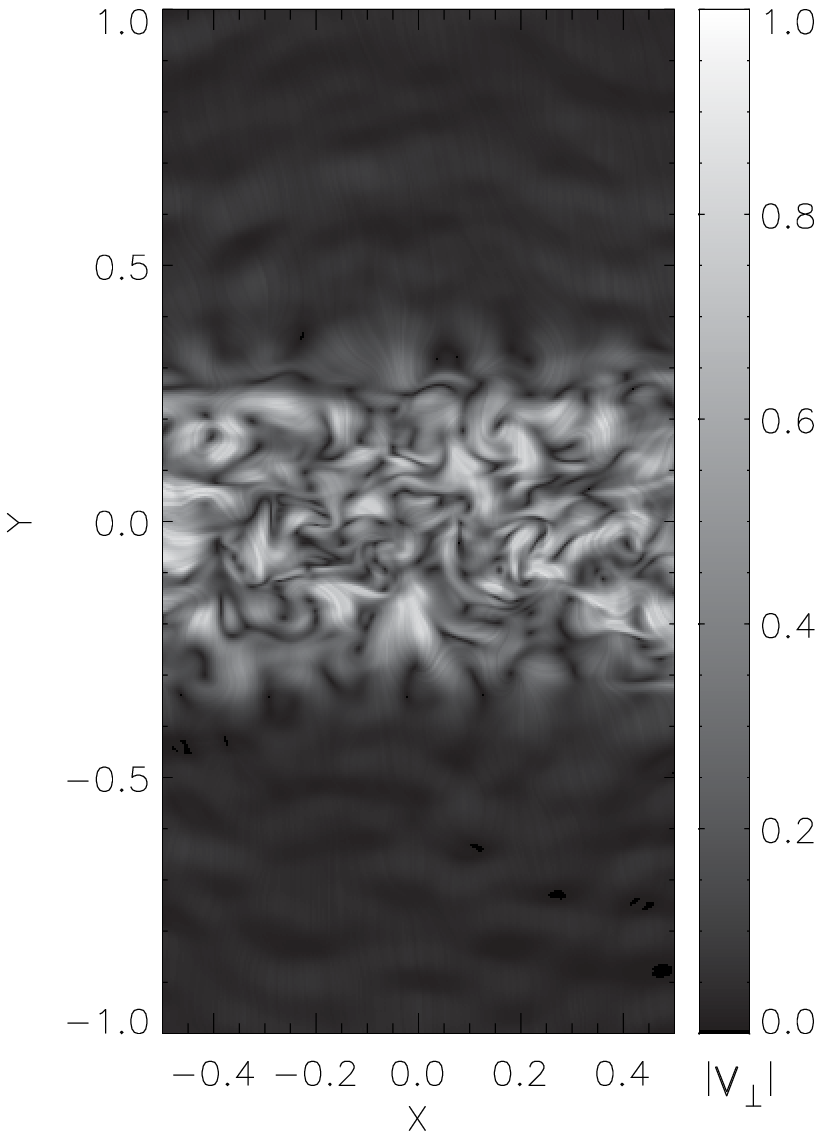}
\includegraphics[width=0.6\columnwidth]{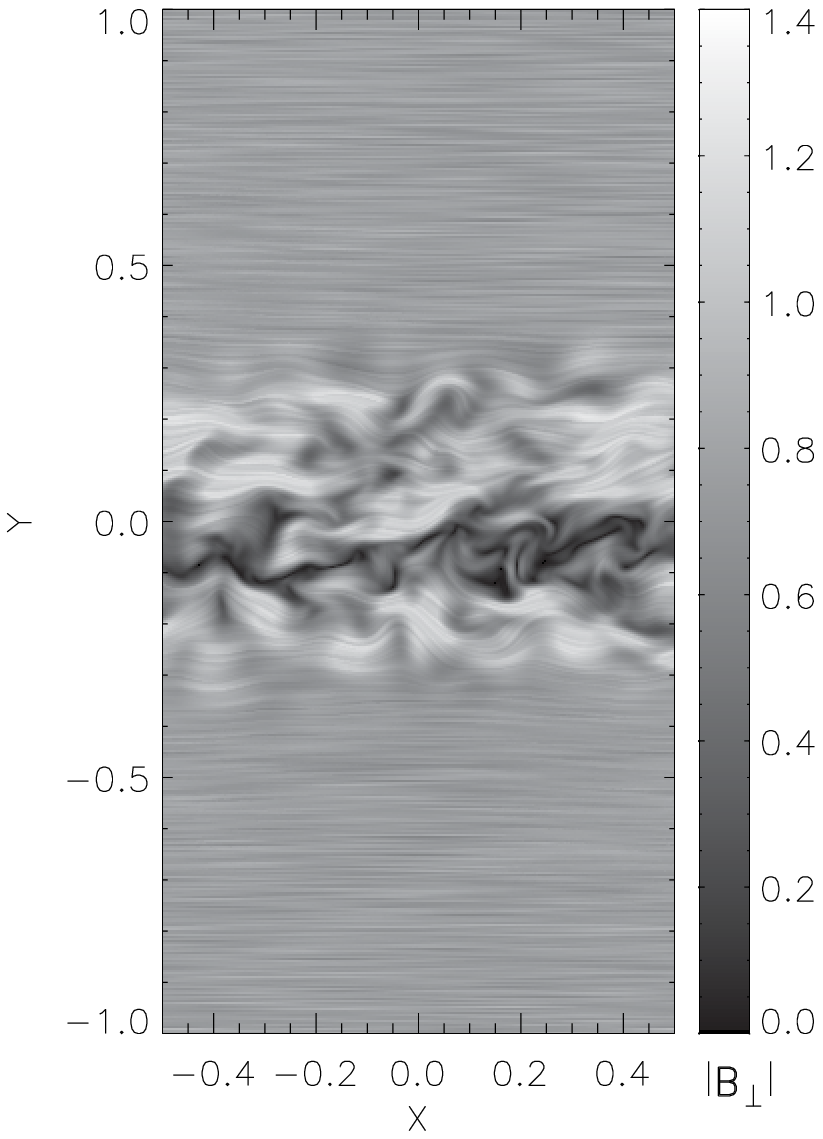}
\includegraphics[width=0.6\columnwidth]{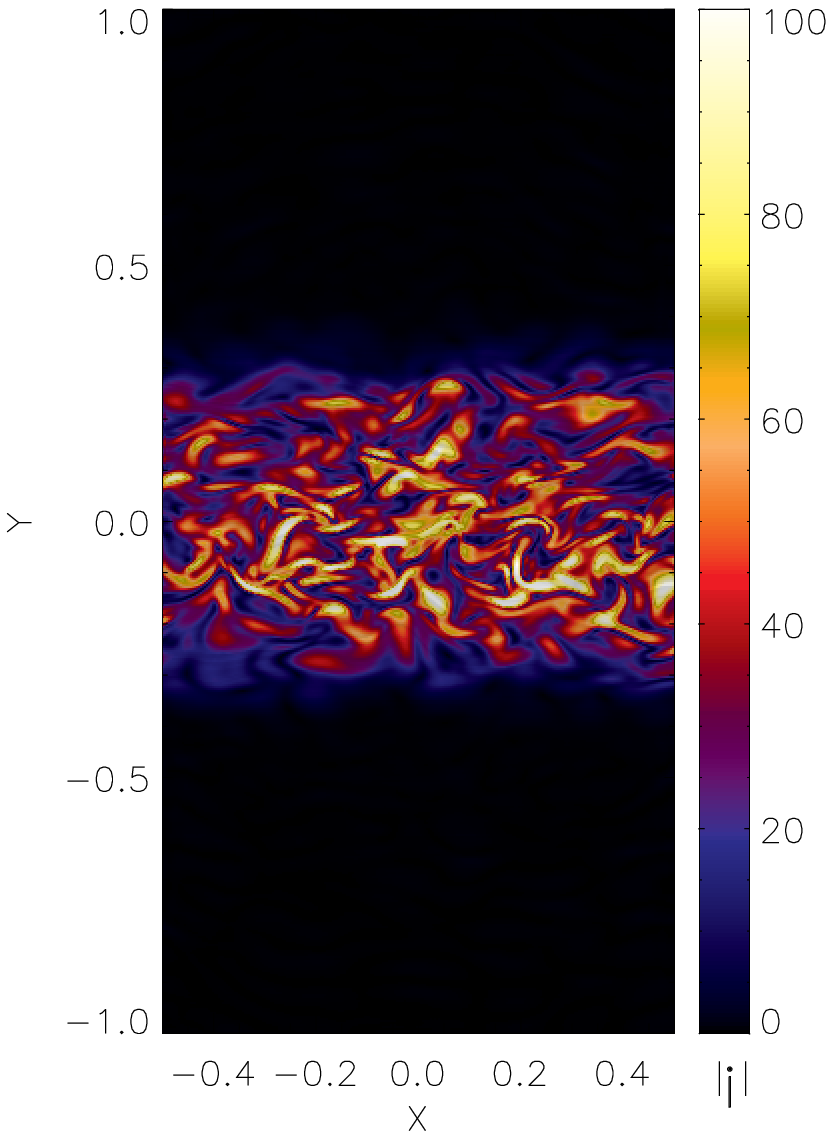}
\includegraphics[width=0.6\columnwidth]{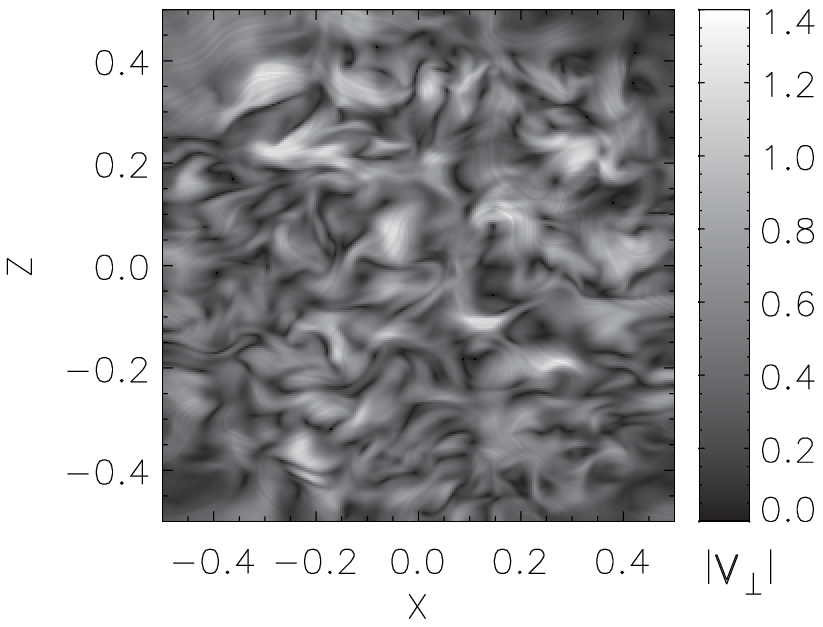}
\includegraphics[width=0.6\columnwidth]{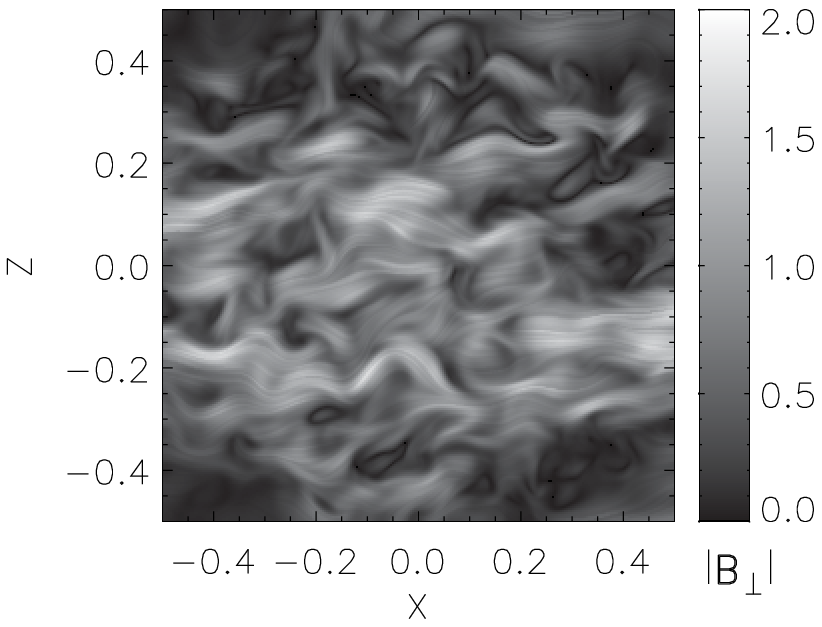}
\includegraphics[width=0.6\columnwidth]{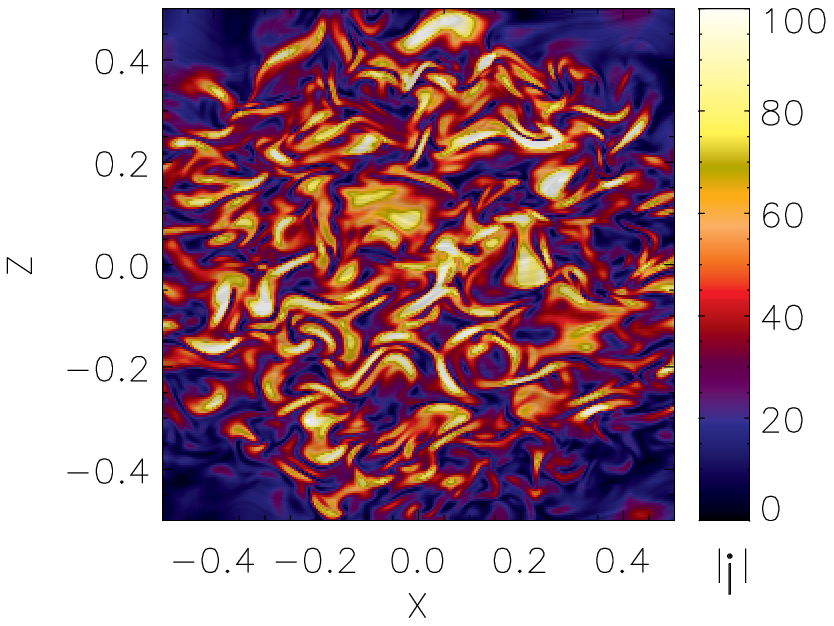}
\caption{Topology and strength of velocity field (left panel) and magnetic field
(middle panel) in the presence of fully developed turbulence at time $t=12$. In
the right panel we show distribution of the absolute value of current density
$|\vec{J}|$. The images show the XY-cut (upper row) and XZ-cut (lower row) of
the domain at the midplane of the computational box. Turbulence is injected with the power $P_{inj}=1$ at scale
$k_{inj}=8$. \label{fig:top_turb}}
\end{figure*}
In Figure~\ref{fig:top_turb} we show examples of XY-cuts (upper row) and XZ-cuts
(lower row) through the box of the velocity (left panel) and magnetic field
(middle panel) topologies with the intensities corresponding respectively to the
amplitude of perpendicular components of velocity and magnetic field to the
normal vector defining the plotted plane.

The first noticeable difference compared to the Sweet-Parker configuration is a
significant change of the velocity and magnetic field topologies. Velocity has
very complex and mixed structure near the mid plane, since we constantly inject
turbulence in this region (see the left panel in Fig.~\ref{fig:top_turb}).
Although the structure is very complex here, most of the velocity
fluctuations are pointed in the directions perpendicular to the mean magnetic
field. This comes from the fact, that in the nearly incompressible regime of
turbulence, most of the fluctuations propagate as Alfv\'en waves along the mean
magnetic field. Slow and fast waves, whose strengths are significantly reduced,
are allowed to propagate also in directions perpendicular to the mean field. As
a result most of the turbulent kinetic energy leaves the box along the magnetic
lines. The fluctuations, however, efficiently bend magnetic lines near the
diffusion region. There, the strength of the magnetic field is reduced, since this
is the place where magnetic lines change their directions (see the middle upper
panel in Fig.~\ref{fig:top_turb}). The interface between positively and
negatively directed magnetic lines is much more complex then in the case of
Sweet-Parker reconnection. This complexity favors creation of enhanced current
density regions, where the local reconnection works faster since the current
density reaches higher values (see the right panel of Fig.~\ref{fig:top_turb}).
Since we observe multiple reconnection events happening at the same time
(compare the right panel of Fig.~\ref{fig:top_turb} to the Sweet-Parker case in
Fig.~\ref{fig:top_sp}), the total reconnection rate should be significantly
enhanced.

%%-----------------------------------------------------------------------------
%%
\subsubsection{Dependence on the Turbulent Power}
\label{ssec:power}

We run several models with varying power of turbulence. All other parameters
were kept the same. This allowed us to estimate the dependence of the
reconnection rate on the power of injected turbulence.

\begin{figure}[t]
\includegraphics[width=\columnwidth]{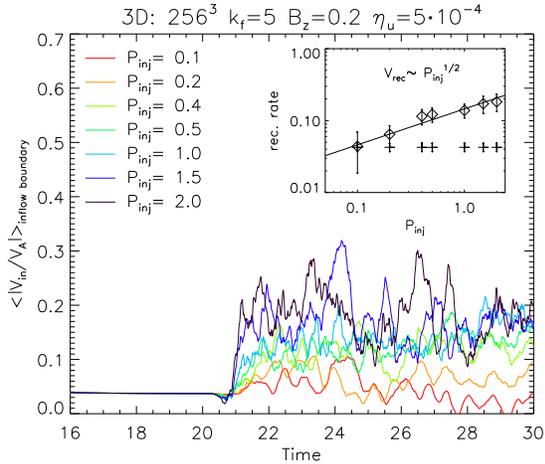}
\caption{Time evolution of the reconnection speed $V_{rec}$ for models with
different powers of turbulence $P_{inj}$ (see legend). In the subplot we show the
dependence of $V_{rec}$ on $P_{inj}$. \label{fig:pow_dep}}
\end{figure}
Figure~\ref{fig:pow_dep} shows the evolution of reconnection speed in models
with the turbulent power $P_{inj}$ varying in the range of over one order of
magnitude, from 0.1 to 2.0. The evolution of $V_{rec}$ reaches stationarity in
a relatively short period of about one Alfv\'en time, estimated from the plot.
In order to obtain the turbulent power dependence, we averaged $V_{rec}$ over a
time interval starting from $t=21$ and ending at time $t=30$. In the subplot of
Figure~\ref{fig:pow_dep} we plot the dependence of the averaged reconnection
speed on the strength of turbulence. Diamonds represent the averaged
reconnection rate in the presence of turbulence. Pluses represent the
reconnection rate during the Sweet-Parker process, i.e. without turbulence. The
error bars correspond to the standard deviation of $V_{rec}$, which is a measure
of time variation.

Fitting to the calculated points gives us a dependency of the reconnection speed
$V_{rec}$ scaling with the power of turbulence as $\sim P_{inj}^{1/2}$. The
power is proportional to $\sim V_l^2/t$ and $t \sim L / V_l \cdot V_A / V_l$,
where $V_l$ is the amplitude of turbulence at the injection scale and $V_A$ is
the Alfv\'en speed. This gives the relation $P_{inj} \sim V_l^4$, thus the
dependency of the reconnection speed $V_{rec}$ on the amplitude of fluctuation
at the injection scale $V_l$ is $V_{rec} \sim V_l^2$, which corresponds to the
LV99 prediction.

%%-----------------------------------------------------------------------------
%%
\subsubsection{Dependence on the Injection Scale}
\label{ssec:injection_scale}

Similar studies have been done in order to derive the dependence of the
reconnection speed $V_{rec}$ on the scale at which we inject turbulence,
$l_{inj}$. Keeping the same power of turbulence for all models we inject
turbulence at several scales, from $k_f = 3$ to $k_f = 12$. This limited range
allows us to obtain desired dependence.

\begin{figure}[t]
\includegraphics[width=\columnwidth]{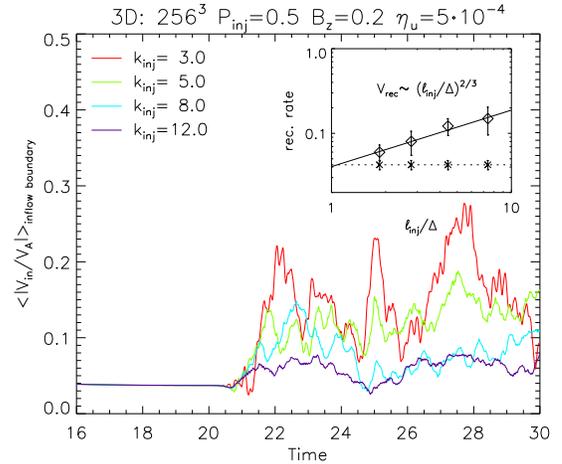}
\caption{Time changes of the reconnection speed $V_{rec}$ for models with
different injection scale $l_{inj}$. In the subplot we show the dependence of
$V_{rec}$ on $l_{inj}$. $\Delta$ is the current sheet thickness, and it is
assumed constant for all models. \label{fig:sca_dep}}
\end{figure}
In Figure~\ref{fig:sca_dep} we present the results obtained in this series of
models. From the plot we clearly see a strong dependence of the reconnection rate
on the injection scale. The model with the injection at a smaller scale $k_f=12$
reaches smaller values of the reconnection rate. In the model with the injection at larger
scale $k_f=3$, the reconnection is faster than the Sweet-Parker reconnection by
a factor of almost 10 at some moments. After averaging the rates over time, we
plot its dependence on the injection scale in the subplot of
Figure~\ref{fig:sca_dep}.

The fitting to the relation of reconnection rate on the injection scale gives
the dependency of $V_{rec} \sim l_{inj}^{2/3}$, which is stronger then predicted
$\sim l_{inj}^{1/2}$ by LV99. In their paper, the authors considered
Goldreich-Sridhar model of turbulence \citep{goldreich95} starting at $l_{inj}$.
The existence of the inverse cascade can modify the effective $l_{inj}$. In
addition, reconnection can also modify the characteristics of turbulence. This
aspect requires more study.

%%-----------------------------------------------------------------------------
%%
\subsubsection{Dependence on Resistivity}
\label{sec:resistivity_dep}

In the global constraint of reconnection rate derived by LV99 there is no
explicit dependency on the resistivity. In order to study this, we performed
another set of models in which we changed the uniform resistivity $\eta$ only.

\begin{figure}[h]
\includegraphics[width=\columnwidth]{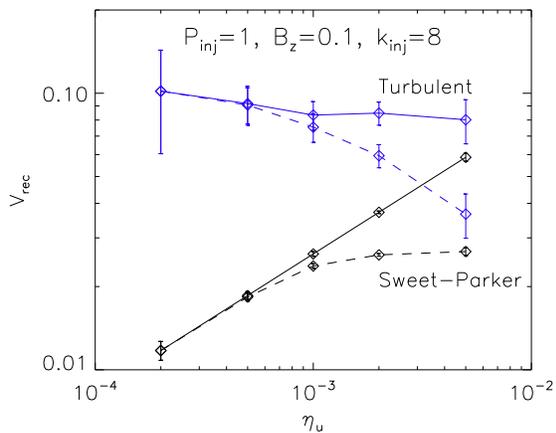}
\caption{Dependence of the reconnection rate on the uniform resistivity. Black
line shows Sweet-Parker reconnection rate, and blue line shows the reconnection
rate in the presence of turbulence. Dashed lines show actual reconnection rates
obtained from models. Solid lines show corrected speeds according to the
procedure described in the text. \label{fig:eta_dep}}
\end{figure}
In Figure~\ref{fig:eta_dep} we show reconnection rates obtained from this set of
models. We plot $V_{rec}$ for five models with $\eta$ varying from $2 \cdot
10^{-4}$ to $5 \cdot 10^{-3}$. The dashed lines show relations obtained directly
from simulations. Since our computational box does not change from model to
model, the magnetic field has non-zero gradients at the boundaries for the
higher values of resistivity. This affects the evolution by reducing the
reconnection rate. We know from theory that the reconnection rate $V_{SP}$
during the Sweet-Parker stage scales as $\sim \eta^{1/2}$. Using this relation,
we have calculated correction coefficients by taking the ratio of
$\eta^{1/2}/V_{SP}$. Using these coefficients we correct values of reconnection
rate obtained during the turbulent stage. The result is plotted with the solid
blue line. The relation signifies virtually no dependence of the $V_{rec}$ on
$\eta$. The correctness of this approach requires more study.

\begin{figure}[t]
\includegraphics[width=\columnwidth]{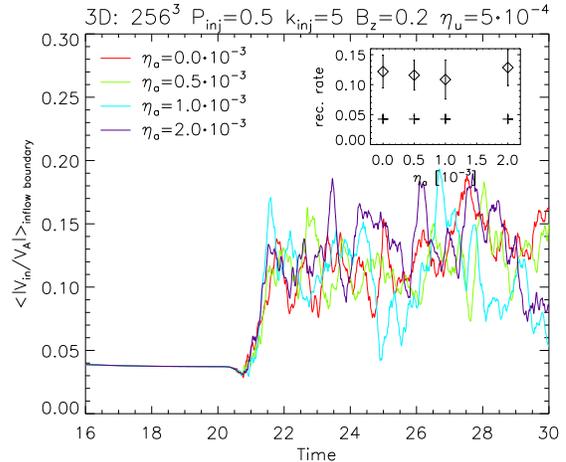}
\caption{The evolution of the reconnection speed $V_{rec}$ for models with
different anomalous resistivity $\eta_a$. In the subplot we show the dependence
of
$V_{rec}$ on $\eta_a$. Critical value of current density $j_{crit}$ is set to
25.0 in all models. \label{fig:aeta_dep}}
\end{figure}
In addition to the uniform resistivity dependence, we have studied the
dependence on the anomalous effects as well. The results of these studies are
presented in Figure~\ref{fig:aeta_dep} and show four models with the same
uniform resistivity $\eta_u = 5 \cdot 10^{-4}$ and the critical current density
$j_{crit} = 25.0$ but with the anomalous resistivity parameter $\eta_a$ varying
between $0.0$ and $2 \cdot 10^{-3}$. In the subplot of Figure~\ref{fig:aeta_dep}
we plot the dependence of the reconnection rate on the anomalous resistivity
parameter $\eta_a$. We see that the reconnection speed $V_{rec}$ is insensitive
to the value of $\eta_a$ to within the variations of the reconnection rate in each model
 (see the error bars).

%%=============================================================================
%%
%%
\section{Discussion}
\label{sec:discussion}

In this paper we tested the model of reconnection proposed by LV99. Our results
show the significant influence of the turbulence on the magnetic reconnection
rate. The reconnection speed $V_{rec}$ (Eq.~\ref{eq:constraint}) shows
dependence on the characteristics of the turbulence only, the rate of energy
injection and the scale of injection. In particular, there is no explicit
dependency on the resistivity.

Numerical testing of the reconnection model in LV99 is far from trivial. The
model is intrinsically three dimensional. In order to develop a sufficient
range of turbulent cascade before it reaches the scale of current sheet we have
to use high resolution simulations. Higher resolution, in addition, minimizes
the role of numerical diffusion. Another problem is the adaptation of proper
boundary conditions. Our model requires open boundaries in order to allow the
ejection of the reconnected flux. This type of boundary conditions should allow
for free outflow of the matter and magnetic field. This property is crucial for
the global reconnection constraint, since the reconnection stops when the
outflow of reconnected flux is blocked.

Even though these numerical simulations allow us to study reconnection in the
presence of turbulence for a limited range of magnetic Reynolds numbers (in this
paper $R_m < 10^3$), the results provide good testing of the relations derived
by LV99. The strong dependence of $V_{rec}$ on the injection scale, when scaled to
the real conditions of the interstellar medium, shows dramatic enhancement of
the reconnection speed, which even in the presence of a magnetic field almost
perfectly frozen in the medium, allows for fast reconnection with the
characteristic time comparable to the Alfv\'en time. The LV99 model predicts this
dependence to be as $V_{rec} \sim l_{inj}^{1/2}$. However, our numerical
testings show a stronger dependence, $V_{rec} \sim l_{inj}^{2/3}$. This
difference can be explained by the presence of the inverse cascade of turbulence
in our numerical models.

The advantage of LV99 model is that it is robust  and fast in any type of fluid,
under the assumption that the fluid is turbulent. Consequences of this are
dramatic. The reconnection process is not determined by Ohmic diffusion,
but it is controlled by the characteristics of turbulence, like its
strength and the energy injection scale. Turbulence efficiently removes the reconnected
flux fulfilling the global constraint (Eq.~\ref{eq:constraint}). Our numerical
results confirm the LV99 prediction of a strong dependence of the reconnection
rate on the amplitude of turbulence, i.e. $V_{rec} \sim V_l^2$, but most
importantly, they show that the reconnection in the presence of turbulence is
not sensitive to the magnetic diffusivity of the medium.

% The reconnection in the presence of turbulence suits very well for number of astrophysical processes. For example, complex and small scale structure of magnetic lines near the diffusion region could efficiently accelerate particles, producing highly energetic cosmic rays \citep[see][]{dalpino01,dalpino05}. Another example is the efficient smoothing of the small-scale into large-scale component of the galactic magnetic field produced by the dynamo, where turbulence are supported by the supernova explosions \citep{hanasz04}. The stochastic reconnection could also explain induction of solar flares.

% The successful numerical testing of the turbulent reconnection model presented in this paper appeal for more studies in this direction. The problems for future studies cover the importance of compressibility in this type of reconnection, since large fraction of observed interstellar medium seems to be supersonic. Also interesting question is the ability of turbulent reconnection to self-sustaining or to generate turbulence by itself. These problems we will address in the next studies.

%%=============================================================================
%%
%%
\section{Summary}
\label{sec:summary}

In this article we investigated the influence of turbulence on the reconnection
process using numerical experiments. This work is the first attempt to test
numerically the model presented by \citet{lazarian99}. We analyzed the
dependence of the reconnection process on the two main properties of turbulence,
namely, its power and the injection scale. We also analyzed the role of Ohmic
resistivity in the weakly turbulent reconnection. We found that:

\begin{itemize}
\item Numerical studies of stochastic reconnection are finally possible, even
though reconnection in numerical simulations is always fast.

\item Turbulence drastically changes the topology of magnetic field near the
interface of oppositely directed magnetic field lines. These changes include the
fragmentation of the current sheet which favors multiple simultaneous
reconnection events.

% \item Fragmentation of current sheet decreases the disparity in inflow/outflow ratios. It means that the reconnection rate does not depend on the Ohmic scales like the thickness of current sheet, but it depends on the scales introduced by turbulence.

\item The reconnection rate is determined by the thickness of the outflow
region. For large scale turbulence, the reconnection rate depends on the
amplitude of fluctuations and injection scale as $V_{rec} \sim V_l^2$ and
$V_{rec} \sim l_{inj}^{2/3}$, respectively.

\item Reconnection in the presence of turbulence is not sensitive to Ohmic
resistivity. The introduction of the anomalous resistivity does not change the
rate of reconnection of weakly stochastic field either.
\end{itemize}

\acknowledgments
The research of GK and AL is supported by the Center for Magnetic
Self-Organization in Laboratory and Astrophysical Plasmas and NSF Grant
AST-0808118. The work of ETV is supported by the National Science and
Engineering Research Council of Canada.  Part of this work was made possible by
the facilities of the Shared Hierarchical Academic Research Computing Network
(SHARCNET:www.sharcnet.ca). This research also was supported in part by the
National Science Foundation through TeraGrid resources provided by Texas
Advanced Computing Center (TACC:www.tacc.utexas.edu).


\begin{thebibliography}
\bibitem[Alvelius(1999)]{alvelius99} Alvelius, K., 1999, Physics of Fluids, 11, 1880
% \bibitem[Beck(2002)]{beck02} Beck, R., {\em Disks of Galaxies: Kinematics, Dynamics and Perturbations}, ASP Conference Proceedings, edited by E. Athanassoula, A. Bosma, and R. Mujica, Astronomical Society of the Pacific, 2002, pp. 331-342
% \bibitem[Biskamp(1996)]{biskamp96} Biskamp, D. \ 1996, \apss, 242, 165
% \bibitem[Biskamp(2000)]{biskamp00} Biskamp, D., {\em Magnetic Reconnection in Plasmas}, Cambridge University Press, 2000
% \bibitem[Cox(2004)]{cox04} Cox, D.~P. \ 2004, \apss, 289, 469
\bibitem[Del Zanna, Bucciantini \& Londrillo(2003)]{delzanna03} Del Zanna, L., Bucciantini, N. \& Londrillo, P., 2003, \aap, 400, 397
\bibitem[Goldreich \& Sridhar(1995)]{goldreich95} Goldreich, P. \& Sridhar, S. \ 1995, \apj 438, 763
% \bibitem[de Gouveia dal Pino \& Lazarian(2001)]{dalpino01} de Gouveia dal Pino, E. \& Lazarian, A. \ 2001, \apj, 560, 358
% \bibitem[de Gouveia dal Pino \& Lazarian(2005)]{dalpino05} de Gouveia dal Pino, E. \& Lazarian, A. \ 2005, A\&A, 441, 845
% \bibitem[Hanasz et al.(2004)]{hanasz04} Hanasz, M., Kowal, G., Otmianowska-Mazur, K., \& Lesch, H. \ 2004, \apj, 605, L33
% \bibitem[Isobe et al.(2006)]{isobe06} Isobe H., Miyagoshi T., Shibata K., \& Yokoyama T. \ 2006, \pasj 58, 423
\bibitem[Lazarian \& Vishniac(1999)]{lazarian99} Lazarian, A., Vishniac, E.~T. \ 1999, \apj, 512, 700, (LV99)
% \bibitem[Lazarian \& Vishniac(2000)]{lazarian00} Lazarian, A., Vishniac, E.~T. \ 2000, Revista Mexicana de Astronom\'\i{}a y Astrof\'\i{}sica Conference Series, 9, 55
\bibitem[Lazarian \& Vishniac(2008)]{lazarian08} Lazarian, A., Vishniac, E.~T. \ 2008, Revista
Mexicana de Astronom\'\i{}a y Astrof\'\i{}sica Conference Series
\bibitem[Moffat(1978)]{moffat78} Moffat, H.~K., {\em Magnetic Field Generation in Electrically conducting Fluids}, Cambridge University Press, London/New York, 1978
% \bibitem[Parker(1957)]{parker57} Parker, E.~N. \ 1957, \jgr, 62, 509
% \bibitem[Parker(1992)]{parker92} Parker, E.~N. \ 1992, \apj, 401, 137
% \bibitem[Petschek(1964)]{petschek64} Petschek, H.~E. \ 1964, {\em Physics of Solar Flares}, AAS-NASA Symposium, NASA SP-50, edited by W. H. Hess, Greenbelt, MD
\bibitem[Priest \& Forbes(2000)]{priest00} Priest, E., Forbes, T., {\em Magnetic Reconnection}, Cambridge University Press, 2000
% \bibitem[Shibata(1999)]{shibata99} Shibata, K. \ 1999, \apss, 264, 129
% \bibitem[Sweet(1958)]{sweet58} Sweet, P.~A., {\em Electromagnetic Phenomena in Cosmical Physics}, Proceedings from IAU Symposium no. 6, edited by Bo Lehnert, Cambridge University Press, 1958
\bibitem[T\'oth(2000)]{toth00} T\'oth, G., 2000, JCoPh, 161, 605
% \bibitem[Vishniac et al.(2002)]{vishniac02} Vishniac, E.~T., Cho, J., \& Lazarian, A. \ 2002, {\em APS Meeting Abstracts}

\end{thebibliography}
\end{document}